# Rethinking meaning and ontologies from the perspective of ontological units

Paul Fabry[a], Adrien Barton[b,a] and Jean-François Ethier[a,1]
[a] *Groupe de Recherche Interdisciplinaire en Informatique de la Santé (GRIIS), Université de Sherbrooke, Québec, Canada*
*E-mail: paul.fabry@usherbrooke.ca. jf.ethier@usherbrooke.ca*
[b] *Institut de Recherche en Informatique de Toulouse (IRIT), Université de Toulouse & CNRS, France*
*E-mails: adrien.barton@irit.fr*

**Abstract**. Ontologies enable knowledge sharing and interdisciplinary collaboration by providing standardized, structured vocabularies for diverse communities. While logical axioms are a cornerstone of ontology design, natural language elements such as annotations are equally critical for conveying intended meaning and ensuring consistent term usage. This paper explores how meaning is represented in ontologies and how it can be effectively represented and communicated, addressing challenges such as indeterminacy of reference and meaning holism. To this end, instead of following the conventional approach of beginning with existing ontologies and working toward alignment or modularization, this article proposes a reversal of perspective: taking the ontological term as the starting point and introducing a new structure, named "ontological unit", characterized by: a term-centered design; enhanced characterization of both formal and natural language statements; and an operationalizable definition of communicated meaning based on general assertions. By formalizing the meaning of ontological units, this work seeks to enhance the semantic robustness of terms, improving their clarity and accessibility across domains. Furthermore, it may offer a more effective foundation for ontology generation and significantly improve support for key maintenance tasks such as reuse and versioning. This article aims to establish the theoretical groundwork for the proposed approach and to lay the foundations for future applications in applied ontologies.

## 1. Introduction

Interoperability, the ability to exchange data and knowledge across diverse domains and systems, is a field of research that has received considerable attention over the years. However, achieving interoperability becomes increasingly challenging with the advent of big data (volume, velocity, variety) and fat data (wide arrays of measurements along varied axes for a given phenomenon). For example, in the health domain, new analytical methods for learning health systems rely on clinical data obtained from patient charts, sensor data, biobank data points, quantified self-time series, environmental data points and social care narratives.

The use of applied ontologies to improve interoperability between information systems has been studied for a while in various domains (Ganzha et al., 2017; Karan et al., 2016; Liyanage et al., 2015). Ontologies offer formal, source-independent representations that are not dependent on specific data models tied to particular technologies or formats. Indeed, a common use of ontologies is to provide a publicly accessible, coherent body of knowledge that supports scientific projects undertaken by various teams worldwide, for example with the OBO Foundry in the biomedical domain (Smith et al., 2007). However, their practical applications assume that a pre-existing, consensual representation of a portion of reality is shared between the ontology's authors and its users, with the ontology's function being to unambiguously refer to specific elements pointed by this shared representation. As suggested in Neuhaus & Hastings (2022), such consensus does not necessarily pre-exist; rather, ontology development involves actively constructing this consensus through a process of mediation and negotiation.

At the heart of the challenges posed by interoperability lies the question of a term's meaning: how can an author specify and communicate the intended meaning of their ontological terms in a way that ensures that it is optimally understood by all users including those that were not part of this consensus building process? In fact, while initially raised in a context of interoperability, this question arises in any situation involving communication between the author of a term and its intended audience.

Furthermore, while an author may regard the intended meaning of a term as invariant, the way it is expressed—through logical axioms and natural language definitions in an ontological artefact—can change over time. As knowledge and requirements evolve, ontological artefacts should therefore be understood as dynamic entities, subject to ongoing modification. To support this, they should incorporate versioning mechanisms that allow existing projects to continue using established artefacts while enabling others to adopt updated versions. Effective versioning, however, depends on the ability to distinguish between a term's meaning characteristics that define its identity, and those that can be modified without compromising its identity, thus enabling the creation of distinct versions of the same artefact.

This work aims to address these questions by examining ontological terms and the determinants of their meaning. Rather than following the conventional approach of starting with existing ontologies and working toward alignment or modularization, this article proposes a reversal of perspective as we take the ontological term as the starting point.

The next sections delve into the characterization of meaning within ontologies, explore how meaning is communicated and highlight current challenges in this area. We then propose strategies to address these challenges and present an approach to better specify the meaning of ontological terms through the use of "ontological units" whose key characteristics are discussed. Such artefact operates upstream of ontologies. Ontological units serve as precursors that can be used in the generation of ontologies. The primary objective of this article is to establish the theoretical foundations for ontological units and the specification of their meaning. A secondary aim is to support their practical implementation within the context of applied ontologies. Accordingly, we concentrate on existing computable solutions, with a particular focus on the OWL 2 language (Cuenca Grau, Horrocks, Motik, et al., 2008). Ontological units can be instrumental in enabling efficient reuse and versioning of ontological terms, and the current article aims at introducing the bases that will be used in that respect in future works.

## 2. Meaning and ontologies

Addressing the question of meaning within ontologies first requires clarifying key terms in our context. The term "ontology" is subject to various interpretations within the field of knowledge engineering, some of which are detailed below. While some of the formalizations discussed here are grounded in first-order logic, we focus on the practical aspects of our work on their applicability to description logic, specifically SROIQ (Horrocks et al., 2006) as implemented in OWL 2 (Cuenca Grau, Horrocks, Motik, et al., 2008).

The notion of meaning has been extensively discussed in philosophy. Among its foundational aspects is Frege's distinction between "sense" and "reference" (Zalta, 2024). According to this distinction, the reference of a term is the actual portion of reality to which the term points. In contrast, the sense of a term concerns the way in which it points to that portion of reality. A classic illustration of this distinction is found in the terms "evening star" and "morning star", which both refer to the same object (the planet Venus) but differ in their senses. Sense and reference are closely aligned with the notion of intension and extension respectively, which are more commonly used in logic and set theory. The extension of a term refers to the set of all entities in the world to which the term applies, while its intension denotes a set of properties or attributes shared by all those entities. Intensions might be represented as a function from possible worlds to extensions, and can be seen as a way of formalizing Frege's idea that sense determines reference (Fitting, 2022).

In this paper, we adopt a classical, so-called internalist view of meaning, which holds that the meaning of a term is determined by internal, cognitive factors rather than external referents. This perspective closely associates meaning with intension, situating it either within the cognitive structures of its users or within the language in which the term is used, both of which are examined below. We also examine how meaning is conveyed and the challenges encountered in its communication.

### *2.1. Meaning within the cognitive structure*

#### *2.1.1. Extensional approach*

The most cited definition of an ontology emanates from Gruber (1993), was later expanded by Borst et al. (1997) and then synthesized by Studer et al. (1998) as follows:

*"An ontology is a formal, explicit specification of a shared conceptualization."*

The terms "explicit" and "formal" indicate that the types of concepts used and the constraints on their application are clearly defined in a formal language, such as first-order logic, as used by these authors. Describing the conceptualization as "shared" means that an ontology captures consensual knowledge, knowledge that is not unique to an individual but is accepted by a group.

The authors draw on the notion of conceptualization defined by Genesereth and Nilsson (1987). A conceptualization consists of a tuple ($S = (D, \mathbf{R})$), composed of a set of objects considered for a given representation, known as the universe of discourse ($D$), along with a set of relations of arity $n$ between these objects ($\mathbf{R}$).

There are several key points to note in this definition of conceptualization. First, Genesereth and Nilsson adopt a broad definition of what is an object. Objects include instances and classes, tangible (such as a physical entity), abstract (like a mathematical concept), or even fictional (such as a unicorn). For these authors, an object is "[…] anything about which we want to say something." (Genesereth & Nilsson, 1987). This conception of the universe of discourse has the benefit of enabling us to discuss hypothetical entities that might not exist in our world, such as a future manned mission to Mars, a hypothetical elementary particle, or even unicorns (Schulz et al., 2011). In particular, this contradicts the realist approach, in which ontologies represent particulars, universals, or defined classes that exist according to our best scientific knowledge (Arp et al., 2015). Proponents of the realist approach argue that by grounding ontologies in widely accepted scientific theories, it provides a common foundation that helps reduce conceptual idiosyncrasies and improves interoperability.

Second, this definition considers a conceptualization as purely extensional, equating a concept solely with its reference. As a result, any change in reference, such as adding or removing an instance for example, alters the conceptualization. This approach ties a conceptualization to a particular state of the world and fails to capture the term's meaning, its intension, which is distinct from its extension.

### *2.1.2. Intensional approach*

To overcome these limitations, Guarino et al. (2009) propose a new definition of an ontology that endorses an intensional approach to conceptualization. The authors expand on Genesereth and Nilsson's conceptualization (the tuple $S = (D, \mathbf{R})$), which they rename as an "extensional relational structure." They define an extensional interpretation function $I$ that maps a language **L** to this structure, assigning each symbol in the vocabulary **V** of **L** either to an element of the universe of discourse $D$ or to an extensional relation in **R**. The pair ($M = (S, I)$) consisting of the extensional relational structure $S$ and the interpretation function $I$ is named an "extensional first-order structure" and will be referred to as a "model for the language **L**" in the remainder of this article (Figure 1).

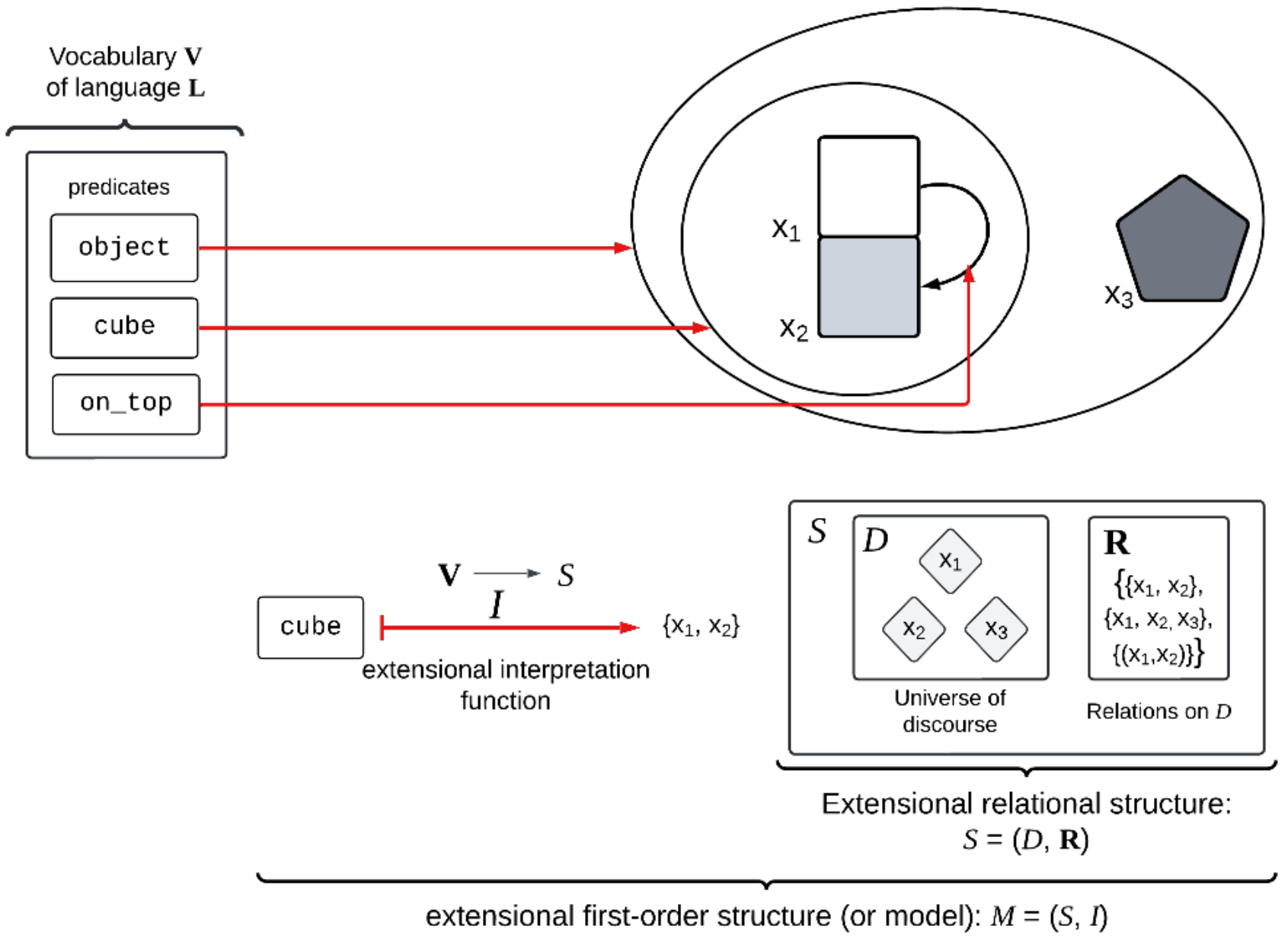

Figure 1 – Representation of an extensional first-order structure (a model for the language L) according to Guarino et al.

The authors base their definition of a conceptualization on the notions of intensions and possible worlds. Within this framework, a concept is identified with an intensional relation $\rho$, that is a function that takes as input a possible world and returns the corresponding extension in that specific possible world. This provides an account of meaning that considers all the different ways the world could have been, or all the possible states of affairs.

Formally, they define a conceptualization, or "intensional relational structure," as a triplet $\mathbf{C} = (D, W, \mathfrak{R})$, where $D$ is a universe of discourse, $W$ is a set of possible worlds and $\mathfrak{R}$ is a set of intensional relations on the domain space $(D, W)$. Each intensional relation $\rho \in \mathfrak{R}$ is a function that takes as input a possible world $w$ belonging to $W$ and returns an element of the universe of discourse ($D$) or an extensional relation belonging to **R**, thus collectively defining an extensional relational structure for every possible world $w$.

The authors also define an "intensional first-order structure," also named an "ontological commitment," as a tuple $\mathbf{K} = (\mathbf{C}, \mathfrak{I})$, where $\mathfrak{I}$ (called an "intensional interpretation function") is a function that associates each vocabulary symbol of $\mathbf{V}$ either with an element of $D$, or with an intensional relation belonging to the set $\mathfrak{R}$.

In summary, while a model directly maps $\mathbf{V}$ to an extensional relational structure in a possible world $w$, an ontological commitment maps $\mathbf{V}$ to an intensional relational structure, which assigns to each possible world of $W$ a corresponding extensional relational structure. The connection between a model and an ontological commitment is established via the notion of "intended model", formally defined by the authors as follows:

A model $M = (S, I)$, with $S = (D, \mathbf{R})$, is called an intended model of $\mathbf{L}$ according to $\mathbf{K}$ iff:

- For all constant symbols $c \in \mathbf{V}$, $I(c) = \mathfrak{I}(c)$
- There exists a world $w \in W$ such that, for each predicate symbol $v \in \mathbf{V}$ there exists an intensional relation $\rho \in \mathfrak{R}$ such that $\mathfrak{I}(v) = \rho$ and $I(v) = \rho(w)$

In other words, an intended model of a language $\mathbf{L}$ according to an ontological commitment $\mathbf{K} = (\mathbf{C}, \mathfrak{I})$ is a model $M = (S, I)$ for which there exists a certain world $w$ such that for every term in the vocabulary $\mathbf{V}$, the interpretation function $I$ assigns to this term the value that the intensional relation $\rho$ associated by $\mathfrak{I}$ to that term takes in that world w (identifying constants with intensional relations that take the same value across all possible worlds).

In this framework, an ontology for an ontological commitment $\mathbf{K}$ is a logical theory in a language $\mathbf{L}$ designed to ensure that its set of models (in the sense of model theory) aligns as closely as possible with its set of intended models (Figure 2).

It is important to recognize that an ontology's specification of a conceptualization is an approximation. To accurately reflect the conceptualization, "as closely as possible" in the words of the authors, it is crucial to select an appropriate vocabulary and domain of discourse. For the authors, this process involves defining a language to describe the conceptualization and constraining its interpretations through axioms that ensure that only models aligned with the intended conceptualization are captured.

Regarding possible worlds, several philosophical theories have been proposed to envision them, with the most prominent being concretism, abstractionism, and combinatorialism (Menzel, 2024):

- Concretism posits that all possible worlds exist concretely and equally. According to this view, there is no fundamental difference between the actual world and other possible worlds. Additionally, an entity does not exist across different worlds (there is no transworld identity) but instead may have counterparts in other worlds.
- Abstractionism, on the other hand, treats possible worlds as abstract entities. The actual world is concrete, whereas possible worlds can be conceived as states of affairs that capture the way things could be. In this conception, entities can exist across various possible worlds (transworld identity is possible).
- Combinatorialism suggests that there is a single concrete actual world, and all possible worlds are constructed by recombining simple elements of the actual world. Transworld identity is also possible in this theory, as possible worlds are recombinations of the same entities. However, it implies that possible worlds can only include entities that already exist in the actual world.

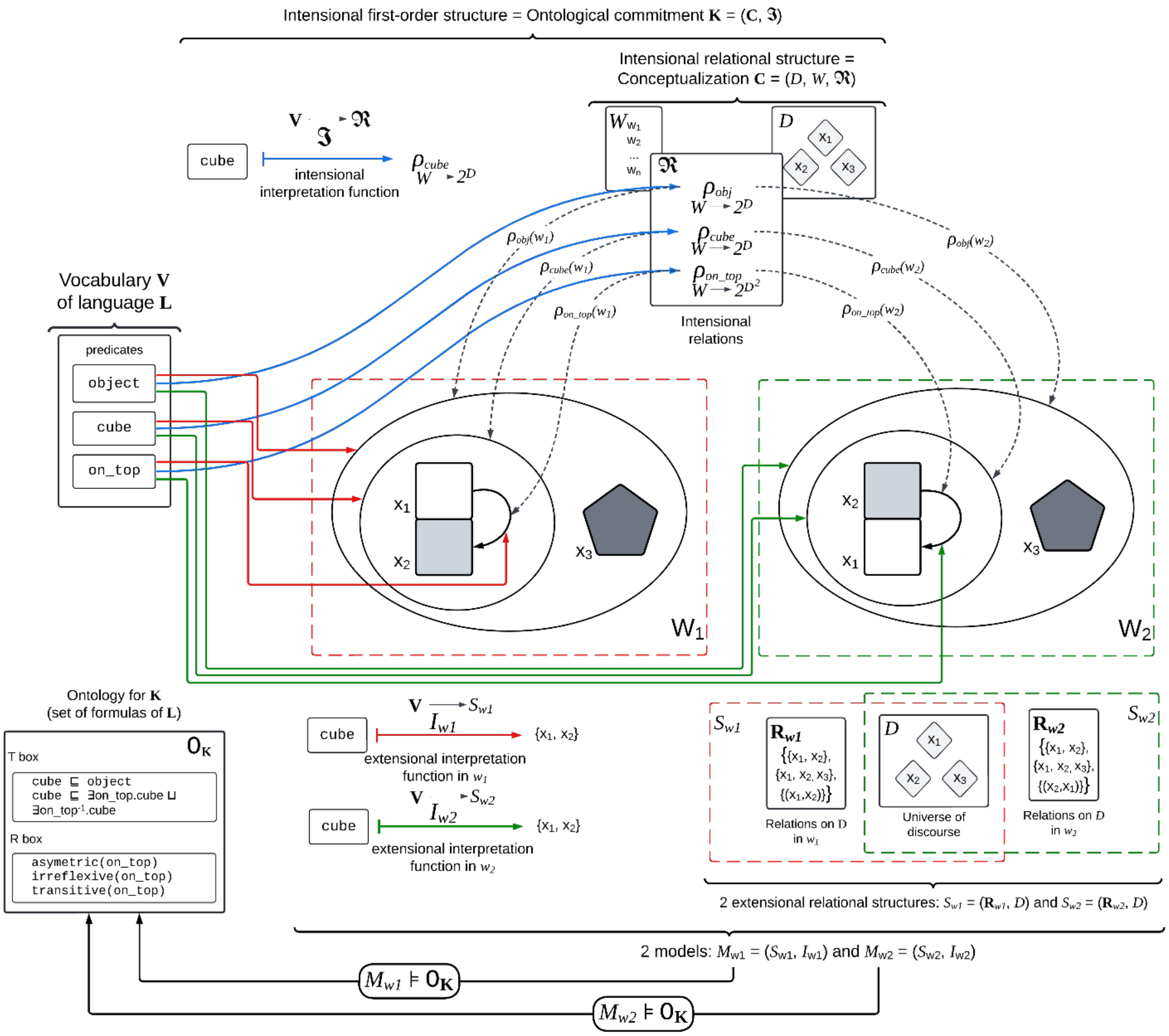

Figure 2 – Representation of an ontology according to Guarino with two models of **V**, $M_{w1}$ and $M_{w2}$, respectively in possible worlds $w_1$ and $w_2$. $M_{w1}$ and $M_{w2}$ are intended models of the ontological commitment **K** because they are compatible with the values taken on by the ontological commitment in their respective worlds. For clarity, it excludes constants from its vocabulary and therefore has an empty A-box.

Guarino et al. propose an interpretation of possible worlds as: "[…] a combination of actual (observed) states of affairs […]" which seems akin to combinatorialism. In that case, all possible worlds are made of the same individuals albeit in different relations.

Some criticisms of this definition of an ontology have been raised. Neuhaus puts forward several arguments, both practical and philosophical (Neuhaus, 2017). Notably, he challenges the proposition that mental conceptualizations are intensional relational structures for several reasons. First, this view would imply that an agent's brain contains information about all possible situations, entities, and the relationships between them—something that is clearly impossible. Second, it would suggest that every conceptualization works as a mapping that associates possible worlds with the extension of that concept in those worlds. This is not necessarily true, as our understanding of certain domains can be superficial and incomplete. As the author suggests as an illustration, we might recognize that birches and aspens are distinct types of trees without being able to distinguish between them in reality, which means we cannot accurately determine their extensions. In addition, as we learn to differentiate between these trees, our conceptualization may evolve without changing the extension itself. Lastly, large ontologies are typically the product of collaborative efforts involving many individuals, each with their own conceptualization of the domain, which contradicts the idea of an ontology representing a singular, shared conceptualization. Instead, Neuhaus suggests that while intensional relational structures are useful formal tools for modelling the human ability to classify objects and relations, they do not reflect the real nature of conceptualizations.

Another important point to note about the previous definitions is that they restrict an ontology to its logical theory, which overlooks much of the information it contains. As a matter of fact, natural language annotations play a crucial role in establishing the authors' intended interpretation and in conveying the full breadth of information within an ontology.

### *2.2. Meaning within the language*

In another article, Neuhaus (2018) emphasizes the importance of vocabulary and annotations in establishing the desired interpretation of an ontology. The author offers the following informal definition: "An ontology for a given domain of interest is a document that provides: 1) a vocabulary to describe the domain, 2) annotations that document and explain the vocabulary, and 3) a logical theory, comprising axioms and definitions, for the vocabulary. Together, these elements enable a competent user to determine the intended interpretation of the ontology."

Indeed, while logical theories upon which ontologies are built may be isomorphic from a purely logical standpoint, their ontological meaning can differ significantly depending on the choice of vocabulary and annotations. The use of natural language terms complements the logical theory, helping users identify the ontology's subject matter, mitigating ambiguities, and understanding the propositions it asserts (Neuhaus & Smith, 2008). To explore these issues, we examine the ontology and universe of discourse shown in Figure 3. For simplicity of exposure, our approach methodologically aligns with Guarino's perspective, in which linguistic symbols points towards intensional relations that, in turn, correspond to specific extensional relations across possible worlds.

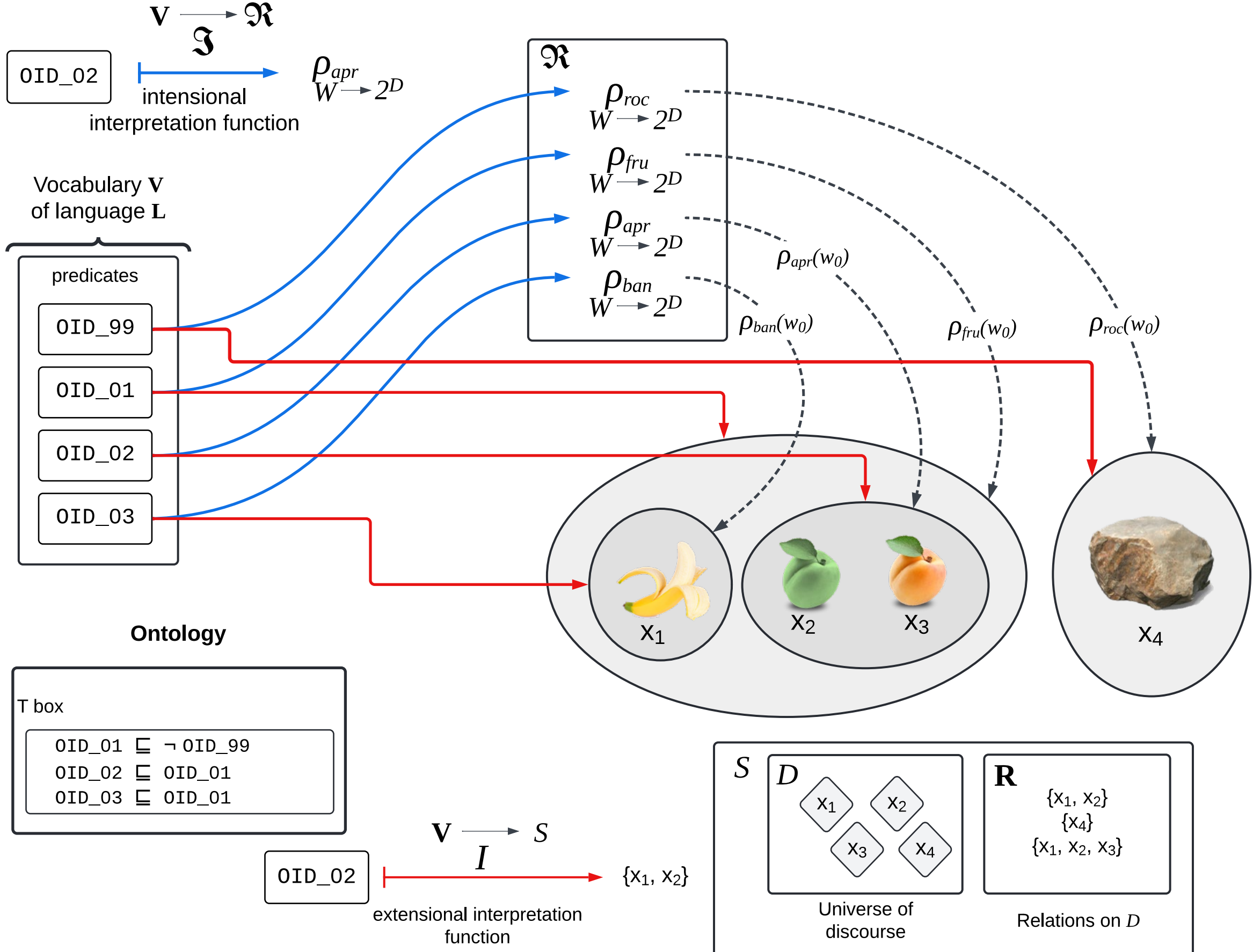

Figure 3 – Example of an ontology and a universe of discourse based on Guarino's framework. For clarity, a single world ($w_0$) is considered, the ontology excludes constants from its vocabulary and encompasses only taxonomical axioms.

### 2.2.1. *Terms and labels*

The ontology vocabulary terms in Figure 3 are globally unique identifiers (GUID), such as "OID_02." Notably, the string "OID" stands for "Ontological IDentifier," a term defined later in this article. These identifiers[1] allow to clearly distinguish the ontology's terms, i.e., the set of non-logical symbols that forms its signature, from the natural language annotations used to interpret them. This reduces ambiguity and facilitates multilingual use. This approach reflects common practice in applied ontologies, where vocabularies often consist of arbitrary terms, such as "http://purl.obolibrary.org/obo/OGMS_0000031" in the OBO Foundry for example.

Let's suppose now that the author adds the label "apricot" to the term "OID_02". Label assignment is often done in applied ontologies through the use of the annotation property "rdfs:label" as follows:

---

[1] For clarity, identifiers like "OID_02" are considered globally unique in this paper. However, in real-world applications, they would contain too few characters to serve this purpose effectively and should be completed by some prefix.

$A_1$: OID_02 rdfs:label "apricot"@en

To note, we adopt RDF's syntax for internationalized strings (RDF 1.1 Concepts and Abstract Syntax, 2014) to specify the language used (e.g., "@en" for English). Notably, the meaning associated with a string varies depending on the language. For instance, the string "coin" constitutes a word referring to a corner in French and another word referring to a round piece of metal used as currency in English.

This annotation property is defined as follows: "rdfs:label is an instance of rdf:Property that may be used to provide a human-readable version of a resource's name." (RDF 1.2 Schema, 2024). This definition clearly distinguishes between a label (e.g., "apricot") and a term (e.g., "OID_02"). According to this distinction, the label "apricot" in $A_1$ would not be interpreted as an intensional relation but rather identify a term (i.e., "OID_02") that does. However, the string "apricot" would constitute on one hand a label identifying a term within the ontology and, on the other hand, for English speakers, a word that points towards an intensional relation which may or may not align with the intensional relation pointed at by "OID_02".

The lack of this distinction in many applied ontologies can lead to confusion. For example, when both a term and its definition are present, users may interpret each differently. In healthcare, labels like "hospitalization" illustrate this issue: users would have differing interpretations of this term (e.g., whether an emergency room stay is part of a hospitalization or not), leading to potential misunderstandings depending on how the term is defined. This also illustrates the importance of definitions.

#### *2.2.2. Definitions*

The primary role of definitions in ontologies is to clarify the meanings of terms and eliminate ambiguity, thereby aligning lexical usage among users (Seppälä, Ruttenberg, Schreiber, et al., 2016). Therefore, ontological definitions serve a stipulative linguistic function by adjusting the recipient's lexical competence to match the usage of proficient speakers within a specific domain. To achieve this, definitions must be carefully crafted to be relevant to their context and target audience. Good practices in writing definitions for ontologies (Seppälä et al., 2017) recommend an Aristotelian structure, often represented by the formula "X is a Y that Z", which consists of three key elements:

- The *definiendum*: This is the term we are trying to define. It is the "X" in the formula.
- The *copula*: This is the element that establishes the link between the *definiendum* and the *definiens*. It is the "is a" in the formula above.
- The *definiens*: This is the set of terms used to define the *definiendum*. It is the "Y that Z" in the formula. The *definiens* is usually composed of two parts:
  - The *genus* (Y): This is the generic term that places the *definiendum* in a broader category.
  - The *differentia*(s) (Z): This is the specific characteristic that distinguishes the *definiendum* from other entities belonging to the same *genus*.

An important aspect of definitions is the relationship between the *definiendum* and the *definiens*, marked by the *copula*. Traditionally, definitions imply an equivalence relation, where the *definiens* provides necessary and sufficient conditions (NSC) for the *definiendum*. However, definitions can also express other types of conditions. Partial definitions, for instance, specify necessary but not jointly sufficient conditions (NC). Ideally, ontologies should contain only definitions with necessary and sufficient conditions, as their linguistic function is to eliminate ambiguity. However, this ideal is not always achievable, as it can sometimes be out of reach to provide a statement that meets both criteria.

Let's consider the terms "OID_02" and "OID_03" from figure 3 that the ontology author defines respectively as "A fruit of the tree Prunus armeniaca." and "A tropical fruit.".

From a practical standpoint, for example in the OBO Foundry, definitions are typically created using the annotation property "definition" (IAO_0000115) from the Information Artifact Ontology (Smith & Ceusters, 2015) as follows:

$A_2$: OID_02 IAO_0000115 "A fruit of the tree Prunus armeniaca."@en
$A_3$: OID_03 IAO_0000115 "A tropical fruit."@en

This property implies that the condition should be necessary and sufficient; however, it is not always used with this intent and may sometimes indicate partial definitions. For example, in $A_2$ the nominal group "A fruit of the tree Prunus armeniaca" is the *definiens*, following the *genus/differentia* structure with "fruit" as the *genus* and "of the tree Prunus armeniaca" as the *differentia*. In this case, the author specifies the *definiens* as both a necessary and sufficient condition for the *definiendum*. In other words, within the context of the ontology every entity characterized by "OID_02" is a fruit of the tree Prunus armeniaca, and every fruit of the tree Prunus armeniaca is characterized by "OID_02". On the other hand, $A_3$ is to be considered a partial definition as the author specifies the *definiens* as a necessary condition for the *definiendum*: while every entity characterized by "OID_03" is a tropical fruit, not every tropical fruit can be characterized by "OID_03".

#### *2.2.3. Analytic/synthetic distinction*

An ontology includes various types of annotations beyond labels and definitions. Relevant annotations encompass any material that documents the vocabulary or logical axioms, such as natural language definitions, comments, and examples. Another way to categorize them is by using the distinction between synthetic and analytic statements, which can be traced back to Kant (Kant, 1787; Rey, 2023). An analytic statement is one whose truth depends on the meanings of its terms alone. A classic example is the statement "All bachelors are unmarried" that is analytically true because the predicate is used in the definition of the subject, e.g., a bachelor is unmarried by definition. In contrast, a synthetic statement is one whose truth does not depend solely on the meaning of its terms. For example, "All bachelors are happy" is a synthetic statement; its truth cannot be determined from the meanings of its terms alone and instead requires empirical verification.

Kant also proposed the distinction between *a priori* beliefs that are justifiable independently of experience, like "A triangle is a polygon with three corners and three sides", and *a posteriori* beliefs, that need to be empirically verified, like "Water boils at 100°C under standard pressure at sea level".

These two distinctions allow the categorization of statements in four types: analytic *a priori*, analytic *a posteriori*, synthetic *a priori*, and synthetic *a posteriori*. The existence of synthetic *a priori* statements (and to a lesser extent, of analytic *a posteriori* statements) is a matter of debate that has been going on among philosophers (Kripke, 1991) since then, and on which we will not take a stance here. In this paper, we consider only analytic statements that are *a priori* and synthetic statements that are *a posteriori*.

While not explicitly framing it as analytic/synthetic, Neuhaus and Hastings (2022) differentiate between "meaning postulates" that "[…] reflect a terminological choice of the ontology developers, and, thus, contain no empirical assertion about the world" and "falsifiable knowledge", which: "[…] would represent a claim about the world, which could be empirically falsified […]". According to these authors, the truth of the latter depends on the meanings established by the former. Thus, the main task of an ontology is to create a formal vocabulary through meaning postulates (analytic), which can then be used to express empirical statements (synthetic). However, this distinction is usually not made explicit in applied ontologies.

Those insufficient characterizations—whether in labels, definitions, or the analytic or synthetic nature of statements—will impede clear communication of the ontology's meaning, compounding its inherent challenges.

### *2.3. Communication of meaning*

According to the definitions mentioned earlier, a key aspect of an ontology is that it is intended to be shared, typically between its authors and a community of users. To achieve this, unambiguous communication of ontology terms is crucial for users to accurately discern the intended interpretation. Previous work has already addressed several issues in communication between an ontology's authors and its users, such as the possibility of making errors (Fabry et al., 2023). Additionally, we point out two significant challenges in this communication: indeterminacy of reference and meaning holism (Barton et al., 2024).

The indeterminacy of reference is a philosophical problem that highlights the difficulty of determining with certainty to which part of reality a term refers. The canonical example is the "Gavagai" scenario proposed by Quine (2013). Imagine a linguist studying an unknown language. A speaker of that language points to a rabbit and exclaims, "Gavagai!" The linguist cannot determine with certainty whether "Gavagai" refers to the rabbit in its entirety, or to a specific part of the rabbit, or to expressions such as "let's eat this" or "we're lucky" if the speaker is superstitious and associates the sight of a rabbit with good luck.

This problem also arises in the field of ontology development due to several factors. In the process of constructing formal definitions in ontologies, terms are defined using other terms, which in turn may rely on yet further terms for their definitions. Eventually, this chain leads to one or both of the following scenarios (Barton et al., 2024):

- Circularity: where the definition of a term $t_0$ ultimately refers to itself (i.e., $t_0$ is defined using a chain of terms that loops back to $t_0$). Circularity can exacerbate indeterminacy of reference, creating a network of interdependent terms whose meanings remain ambiguous.
- Primitiveness: where certain terms remain undefined, lacking any formal statement within the ontology. Primitive terms, without formal clarification, are open to varying interpretations. Different users may conceptualize them differently, even if they agree on the same ontological statements.

Meaning holism, as traditionally defined (Jackman, 2020), asserts that the meaning of a term depends on the meanings of all other terms within a language. If that ubiquitous, meaning holism would present a significant challenge to ontology development, as any change in the meaning of one term would impact the meanings of all others. Our work has shown that meaning holism presents substantial challenges for ontology engineering which can be mitigated by adopting a certain conception of meaning (Barton et al., 2024) that will be extended below.

Both meaning holism and indeterminacy of reference highlight the delicate nature of meaning communication. From a practical point of view, this challenge is further complicated by the tendency to reuse ontological terms in contexts that may diverge significantly from their current use. This issue is particularly relevant in biomedical ontologies, where reusing existing ontologies or their components is encouraged (Smith et al., 2007). While this practice enhances flexibility in ontology creation and allows for leveraging domain expertise from other groups, there is a risk of altering their intended meaning depending on the axioms imported from other ontologies. This issue is magnified in interdisciplinary contexts, where individuals from different fields are more likely to have significantly divergent conceptualizations associated with similar terms. However, some mitigation strategies can be considered.

## 3. Mitigating meaning communication challenges

Based on our previous work, we propose a multi-pronged approach to address these issues. For exposition reasons we build upon the influential framework proposed by Guarino et al. (2009) and will discuss its limits (especially those by Neuhaus (2017)) in the Discussion section. Specifically, we recommend adopting a term-centered approach, endorsing a formalized specification of meaning, and providing a unified framework for logical and natural language statements.

### *3.1. Term-centered approach*

Traditional ontological engineering often represents an entire domain through a network of terms, with many deriving their meaning from axioms expressing their relationships to one another. Meaning holism and the indeterminacy of reference pose a challenge to the semantic robustness of such a network. Regardless of the network's size or how well defined are the constraints regimenting the use of those terms, their understanding ultimately depends on the choice of primitive terms or textual definitions. While this vulnerability may not be evident when the network is viewed as a whole within its original community, it can become more problematic when part of the network is considered in a different context, such as when integrated into another ontology.

To address this, we propose that each ontological term is created with the aim to identify a single intensional relation, a function that maps each possible world to an extensional relation within that world (see Figure 2). This intensional relation is the term's meaning as intended by the term's author.

Investigating the ontological nature of this intensional relation, whether within the author's cognitive structure or elsewhere, is beyond the scope of this article. However, we assume methodologically an ideal case where the author(s)[2] is fully aware of this intensional relation and aims to communicate it as accurately as possible. To achieve this, it is first essential to be able to refer to this intensional relation unambiguously by using an ontological identifier (OID), defined as follows:

*Ontological identifier=def. A globally unique identifier resulting from a dubbing process during an ontological engineering activity that identifies an intensional relation.*

An important aspect of this definition is that the dubbing process—where a semantic value is arbitrarily associated with a symbol—occurs within ontological engineering activities (Neuhaus & Hastings, 2022). An OID is specifically created to identify an intensional relation (its intended meaning), rather than repurposing an existing identifier for this purpose. To note, for the remainder of this article, the intensional relation identified at by e.g. "OID_01" will be referred to as "$\rho_{\text{OID_01}}$".

The author aims at communicating the term's intended meaning through various statements that describe the intensional relation that a reader may associate to this term. The objective is for users to construct their own intensional relation, referred to here as the "understood meaning," through their interpretation of these statements, aligning it as closely as possible with the author's intended meaning. Based on the above considerations on ontological term's meaning, two characteristics seem especially important for these statements: their analytic or synthetic nature, and their status of necessary, sufficient, or necessary-and-sufficient condition. Additionally, focusing the approach around an OID also requires formulating all relevant statements specifically in relation to that OID. However, logical axioms are usually not intrinsically associated to a particular term. For instance, in description logic an axiom like "OID_02 ⊑ OID_01" is not more intrinsically associated with OID_02 than with OID_01.

[2] In complex ontologies where multiple authors may work on the same term, Neuhaus notes that these reflect diverse, often significantly different, conceptualizations rather than a single shared one. For simplicity, we set aside this criticism in this article.

These statements are referred to as "Ontological Identifier Statements" ("OID statements" for short) because they pertain to a specific OID and are defined as follows:

*Ontological identifier statement=def. A statement that provides a characterization of the intensional relation identified by the OID of interest.*

In order to better illustrate our examples in this article we propose representing them as a quadruplet[3] specifying: a subject OID, the analytic or synthetic nature of the statement, the type of condition it satisfies—necessary (has_NC), sufficient (has_SC), or both necessary and sufficient (has_NSC)—and the characterization, which consists of the natural or formal language elements the OID's author deems suitable for conveying its intended meaning or other relevant information in the case of synthetic statements.

The subject OID and the characterization each point towards an intensional relation that maps an extensional relation to each world, and the condition type defines the relation between these extensional relations for some possible world(s). For example, a characterization being an analytic necessary condition for an OID means that for every world the extensional relations relative to the OID are included in the extensional relations relative to the characterization. This approach extends the classical structure of a definition (definiendum, copula, definiens) for broader applicability. It accommodates statements beyond strict definitions, such as synthetic statements or statements expressing sufficient conditions. In this article, the syntax for writing such statements separates each part with a pipe (|) as follows:

*OID | A/S indicator | condition type | characterization*

To note, the characterization can be another OID, a logical expression using OIDs and logical symbols, or a natural language expression, with each pointing to an intensional relation. For example, the previous axiom "OID_02 ⊑ OID_01" can now be expressed, if we consider it as analytic, as one of the two following OID statements: "OID_02 | Analytic | has_NC | OID_01" and "OID_01 | Analytic | has_SC | OID_02," respectively pertaining to OID_02 and OID_01. Likewise, the previous definition $A_2$ (e.g., OID_02 IAO_0000115 "A fruit of the tree Prunus armeniaca."@en) can be expressed as the following OID statement: "OID_02 | Analytic | has_NSC | "A fruit of the tree Prunus armeniaca."@en"

This unidirectional representation of statements, alongside the analytic/synthetic distinction previously mentioned, enables the formal specification of an OID's meaning.

[3] The quadripartite structure is introduced to more effectively convey the role of these statements; an operationalized structure will be proposed in future work.

### 3.2. *Formalized specification of meaning*

In Barton et al. (2024), we proposed a conception of the meaning of an ontological term grounded in the principle that its meaning is determined by general analytic claims about it, that is, analytic statements expressing a necessary (that may be a necessary and sufficient) condition. Indeed, knowledge representation efforts primarily focus on determining the essential, universally applicable features of a term (Munn & Smith, 2008). To clarify a point of terminology, what was previously referred to as “meaning” in our earlier work corresponds to what we now term “meaning specification.” In the current framework, “meaning” is understood as the “intended meaning”, represented by the intensional relation itself (as in Barton et al., 2025).

In this respect, analytic statements, whether expressed in the ontology’s logical language or in natural language, aim to capture the fundamental aspects of the intensional relation identified by the ontological identifier (OID). Among them, Seppälä et al. (Seppälä, Ruttenberg, & Smith, 2016) emphasizes the primacy of necessary conditions over sufficient ones, as the former provide general characterizations of a term, applying to all instances of that term. Furthermore, statements expressing sufficient conditions can also be interpreted as necessary conditions when viewed in reverse and thus effectively express a general claim on a term possibly outside of the initial scope, carrying the risk of altering its meaning.

Therefore, synthetic statements or statements expressing a sufficient condition on a term are not part of the meaning specification of this term. Additionally, some statements should be excluded from the meaning specification, such as “$\text{OID_02} \sqsubseteq (\text{OID_01} \sqcup \neg\text{OID_01})$”. They are referred to as “logically reducible statements”:

*Logically reducible OID statement =def. An OID statement S1 such that there is another OID statement S2 which has the same subject OID as S1 and a characterization C2 that has the same intension as the characterization C1 of S1, but mentions only OIDs that belong to a subset of the set of OIDs mentioned by C1 and where C2 is a shorter characterization logically equivalent to C1.*

Consider, for example, the following statements[4]:

$A_4$: $\text{OID_02} \sqsubseteq \text{OID_03} \sqcap (\text{OID_01} \sqcup \neg\text{OID_01})$
$A_5$: $\text{OID_02} \sqsubseteq \text{OID_04} \sqcap (\text{OID_04} \sqcup \neg\text{OID_04})$

$A_4$ and $A_5$ are logically equivalent to:

$A_4'$: $\text{OID_02} \sqsubseteq \text{OID_03} \sqcap \top$
$A_5'$: $\text{OID_02} \sqsubseteq \text{OID_04} \sqcap \top$

which in turn are logically equivalent to:

$A_4''$: $\text{OID_02} \sqsubseteq \text{OID_03}$
$A_5''$: $\text{OID_02} \sqsubseteq \text{OID_04}$

---

[4] We thank Fabian Neuhaus and an anonymous reviewer for pointing these examples.

Thus, the right-hand sides of $A_4''$ and $A_5''$ have the same intensions as, respectively, the right-hand sides of $A_4$ and $A_5$ while eliminating the mention to OID_01 in $A_4$, and with a logically equivalent shorter characterization in both $A_4$ and $A_5$.

Only non-logically reducible analytic statements that capture the general aspects of an intension—articulated through statements of either necessary conditions (NC) or necessary and sufficient conditions (NSC)—are included in the meaning specification. These statements are referred to as “meaning-specifying OID statements” and the collection of these statements for a given OID is named “meaning specification” of this OID.

Considering the previous example, “OID_02 | Analytic | has_NC | OID_01” expresses a necessary condition on OID_02 and is thus a member of OID_02’s meaning specification, while “OID_01 | Analytic | has_SC | OID_02” expresses a sufficient condition on OID_01 and is therefore *not* a member of OID_01’s meaning specification.

### *3.3. Unified framework for logical and natural language statements*

Natural language plays a crucial role in ontologies (Neuhaus & Smith, 2008), as they would likely be unintelligible without natural language annotations. In standard ontologies, natural language statements are clearly distinct from logical axioms. As noted earlier, natural language is susceptible to meaning holism and indeterminacy of reference. Even when two individuals agree on an ontology’s formal axioms, their interpretations may differ due to varying understandings of the associated natural language descriptions (Fabry et al., 2023) (Figure 4).

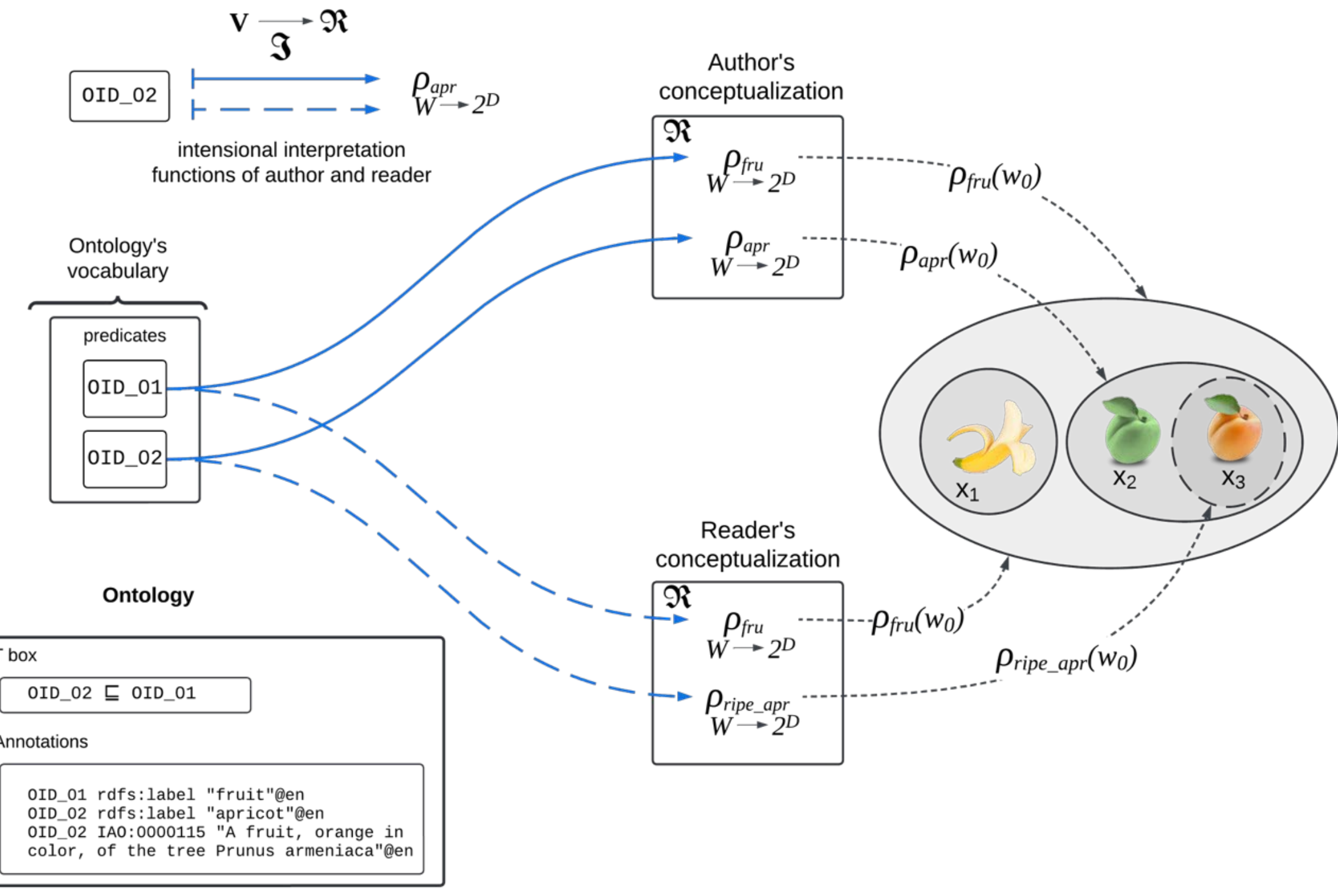

Figure 4 – Representation of a possible misunderstanding between an author and a reader following Guarino's framework. In this example, the reader associates the word "apricot" only to the ripe fruits and assumes that the extension in $w_0$ of the intensional relation identified by "OID_02" is limited to ripe apricots.

Establishing a clear link between formal and natural language statements is essential. Let us now reconsider the definitions mentioned above:

$A_2$: OID_02 IAO_0000115 "A fruit of the tree Prunus armeniaca."@en
$A_3$: OID_03 IAO_0000115 "A tropical fruit."@en

OID statements allow us to express the different types of definitions implicit in natural language. For instance, $A_2$* and $A_3$* explicitly articulate the necessary and sufficient conditions, and the necessary conditions, respectively, as conveyed in $A_2$ and $A_3$:

$A_2$*: OID_02 | Analytic | has_NSC | "A fruit of the tree Prunus armeniaca."@en
$A_3$*: OID_03 | Analytic | has_NC | "A tropical fruit."@en

An important philosophical consideration with these statements lies in the handling of natural language expressions, as each contains multiple distinct terms (e.g., "fruit," "tree," "Prunus armeniaca") that might be interpreted independently. Theories of meaning in natural language are a subject of ongoing debate in philosophy, and a full exploration lies beyond the scope of this article. An important notion in this domain is *compositionality*: the idea that the meaning of a complex expression is determined by the meanings of its parts, or lexical units, and how they are combined. This principle helps explain our ability to understand a possibly infinite number of sentences we have never encountered before (Szabó, 2024). Although there is no universally accepted definition of a lexical unit, it can be understood as an element of a language's lexicon that carries its own meaning and can be combined with other lexical units to form complex expressions. Lexical units are not limited to single words; groups of words, such as idiomatic expressions (e.g. "let the cat out of the bag" or "yellow fever") can also be considered lexical units.

We will adopt the methodological view that the natural language components of such annotations can be considered holistically, as lexical units, serving as input for an interpretation function in their entirety and thereby denoting an intensional relation. In other words, for the author of the OID statement $A_2$* and $A_3$*, "A fruit of the tree Prunus armeniaca." and "OID_02" identify the same intensional relation, while "A tropical fruit." points towards an intensional relation whose extension in each world includes the extension of the intensional relation identified by "OID_03" in this world.

However, not all elements of natural language in an ontology are meant to identify intensional relations. Currently, the rdfs:label annotation property does not differentiate between natural language elements used for their mnemonic value and those that precisely represent the intensional relation they identify. In our view, this also constitutes a significant challenge for ontological alignment, as it carries the risk of alignment based on labels rather than on meanings.

To eliminate this confusion, we introduce the notion of "human-readable tag" that is defined as an information content entity constituted by a string chosen for its mnemonic value to denote an ontological term. Thus, a human-readable tag does not contribute to the meaning specification of an OID and does not pertain to a specific OID. Taking as example $A_1$:

$A_1$: OID_02 rdfs:label "apricot"@en

An author would have at least two (non-exclusive) options. On the one hand, the author could choose to acknowledge the semantic of the word "apricot" (i.e., the intensional relation it identifies) and include it in the ontology's vocabulary with the following OID statement: "OID_02 | Analytic | has_NSC | "apricot"@en". On the other hand, the author could characterize the string "apricot" as a human-readable tag and instead rely on the semantics of the nominal group "A fruit of the tree Prunus armeniaca." to better communicate its intended meaning.

We also consider natural language characterizations to be an integral part of the logical theory's signature, permitting their use within the logical theory alongside OIDs: so, in the above example, "A fruit of the tree prunus armeniaca"@en is part of the logical theory's signature exactly like "OID_01" and "OID_02". Rules for regimenting their use will be detailed in the next section.

To enable the practical application of the strategies for mitigating meaning communication challenges discussed in this section, we introduce the notion of "ontological unit".

## 4. Ontological unit: operationalizing meaning

As mentioned earlier, we methodologically assume the existence of an intensional relation which its holder wants to communicate as accurately as possible to others. This intensional relation is identified by an OID and is communicated through OID statements.  expressed in both natural and formal languages.

This set of statements, categorized in detail below, are manipulated and shared through the use of an ontological unit (OU), an informational entity comprising all OID statements associated with a given OID of interest. Accordingly, each OU is uniquely associated with a single OID. We describe the main components of an OU, then provide a comprehensive definition of an OU's meaning, and discuss how it can be operationalized using the underlying logic.

*4.1. Ontological unit components*

An OU aims to provide all the necessary elements for sharing an intensional relation through a collection of ontological unit statements (OU statements) defined as follows:

*Ontological unit statement=def. A statement that is included in an ontological unit.*

OID statements play a central role among them. A key component of our approach is the analytic/synthetic distinction, which we propose to integrate into Guarino's framework for representing conceptualizations. Specifically, we aim to formalize this distinction within the context of possible worlds semantics. In order to do so, it is important to articulate the analytic/synthetic distinction in relation to the metaphysical distinction between necessary statements (those that hold in all possible worlds) and contingent statements (those that hold in the actual world but not in all possible worlds). In this paper, we adopt the view that analytic statements are metaphysically necessary, whereas synthetic statements are metaphysically contingent. A detailed justification of this position lies beyond the scope of the present article but is provided in a separate work (Barton et al., 2025). This enables us to classify analytic statements into three types as follows:

*Analytic necessary OID statement=def. An OID statement stating that, for every world $w \in W$, the extensional relation associated to w by the intensional relation identified by its subject OID is included in the extensional relation associated to w by the intensional relation pointed at by its characterization.*
Example: OID_02 | Analytic | has_NC | OID_01, which stipulates that for every world $w \in \mathrm{W}$, $\rho_{\mathrm{OID_02}}(\mathrm{w})$ is included in $\rho_{\mathrm{OID_01}}(\mathrm{w})$.

*Analytic sufficient OID statement=def. An OID statement stating that, for every world $w \in W$, the extensional relation associated to w by the intensional relation pointed at by its characterization is included in the extensional relation associated to w by the intensional relation identified by its subject OID.*
Example: OID_99 | Analytical | has_SC | OID_01, which stipulates that for every world $w \in \mathrm{W}$, $\rho_{\mathrm{OID_01}}(\mathrm{w})$ is included in $\rho_{\mathrm{OID_99}}(\mathrm{w})$.

Necessary and sufficient OID can be simply defined as OID statements that are both necessary and sufficient.

*Analytic necessary and sufficient OID statement=def. An OID statement that is both an analytic necessary OID statement and an analytic sufficient OID statement.*

That is, an analytic necessary and sufficient OID statement stipulates that, for every world w ∈ W, the extensional relation associated to w by the intensional relation identified by its subject OID is identical to the extensional relation associated to w by the intensional relation pointed at by its characterization.
Example: OID_02 | Analytical | has_NSC | "A fruit of the tree Prunus armeniaca"@en, which stipulates that for every world $w \in$ W, $\rho_{\text{OID_02}}$(w) is identical with $\rho_{\text{OID_01}}$(w).

In contrast, the truth value of a synthetic statement does not derive solely from the meanings of its constituent terms. Rather, it typically requires empirical verification within the actual world and cannot be assumed to hold across all possible worlds; such statements are considered as metaphysically contingent and are classified as follows:

*Synthetic necessary OID statement=def. An OID statement stating that, at least for our world but not for every world w ∈ W, the extensional relation associated to w by the intensional relation identified by its subject OID is included in the extensional relation associated to w by the intensional relation pointed at by its characterization.*

*Synthetic sufficient OID statement=def. An OID statement stating that, at least for our world but not for every world w ∈ W, the extensional relation associated to w by the intensional relation pointed at by its characterization is included in the extensional relation associated to w by the intensional relation identified by its subject OID.*

*Synthetic necessary and sufficient OID statement=def. An OID statement that is both a synthetic necessary OID statement and a synthetic sufficient OID statement.*

Although OID statements are of great importance, they are not the only type of statement relevant to an OU. Neuhaus (2018) distinguishes between assertive and non-assertive statements:

- Assertive statements include logical axioms and assertive annotations, aiming to assert true propositions about the ontology's domain. The information in assertive annotations can be partially or fully represented by the logical axioms; however, complexities in formalization or limitations of the logical language may prevent a complete representation.
- Non-assertive statements are annotations that do not assert truths about the ontology's domain; rather, they provide contextual information or metadata. These annotations are typically not incorporated into the ontology's logical theory, as they do not directly relate to the modelled domain.

We adopt a similar classification, distinguishing between assertive statements, the OID statements previously mentioned, and non-assertive statements that include the previously mentioned human-readable tags, as well as other types of metadata. These non-assertive statements are referred to as "metadata OU statement" and are defined as follows:

*Metadata OU statement=def. An OU statement that provides a characterization of an OU.*

While an OID statement characterizes an intensional relation, a metadata OU statement characterizes the OU itself. These statements are not further addressed here as they are not directly involved in the meaning specification. They will be detailed in a subsequent work on the implementation of ontological units.

### *4.2. Underlying logic*

We implement ontological units in this paper in description logic (DL), specifically a fragment of the SROIQ language. The implementation addresses two cases: first, the translation of OID statements into DL statements for integration into an ontology; and second, the generation of a meaning specification for an OID based on available OID statements. The latter adds a reverse translation from DL statements back to OID statements.

#### *4.2.1. Translation of OID statements into DL statements*

First, it is important to note that analytic and synthetic statements are treated identically in the DL, as the distinction between them is not expressed in the resulting DL statements. In the examples below, we use the notation “[A/S]” to indicate when the analytic or synthetic nature of an OID statement is irrelevant.

Second, the condition type in OID statements can be expressed in DL according to the equivalences in Table 1. To note, all characterizations are expressed in DL, with natural language expressions regarded as integral components of the signature of the logical theory. Thus, for example, the string ‘*“a mature ovary of a seed-bearing plant”@en’* is to be interpreted not as a descriptive sentence in English but as a class name in DL, on the same pars as *OID_02* for example. Additionally, when the characterization is a DL anonymous class, an OID statement expressing a sufficient condition may be translated into a general class axiom in DL. For example, the following OID Statement: “OID_09 | [A/S] | has_SC | ∃OID_10.OID_11” will be translated in a general class axiom: “∃OID_10.OID_11 ⊑ OID_09”. However, if the characterization is also an OID, it may be represented simply as a taxonomic axiom.

Table 1 - OID statements type condition and their equivalence in description logic axioms.

| **OID statement** | **Description logic axiom** |
|---|---|
| OID_A \| [A/S] \| has_NSC \| characterization | OID_A ≡ characterization |
| OID_A \| [A/S] \| has_NC \| characterization | OID_A ⊑ characterization |
| OID_A \| [A/S] \| has_SC \| characterization | characterization ⊑ OID_A |

#### *4.2.2. Reverse translation of DL statements into OID statements*

As previously noted, the translation of OID statements into DL statements does not preserve[5] the analytic/synthetic distinction. Consequently, a reverse translation cannot recover this information. However, since the reverse translation pertains exclusively to statements that specify meaning, which are analytic, it is assumed that the resulting OID statements are analytic.

[5] While it is possible to envisage implementation strategies that preserve this distinction, using axiom annotations for example, such approaches lie beyond the scope of the present article.

Furthermore, because the language’s signature includes some natural language expression alongside OIDs, a verification step is required during reverse translation to ensure that the OID statements resulting from inferred DL axioms are valid. Consider the following meaning-specifying OID statements:

OS1: OID_01 | Analytic | has_NSC | “A mature ovary of a seed-bearing plant.”@en
OS2: OID_02 | Analytic | has_NSC | “A fruit of the tree Prunus armeniaca.”@en
OS3: OID_02 | Analytic | has_NC | OID_01

These statements are translated into description logic as follows:

DL1: OID_01 $\equiv$ “A mature ovary of a seed-bearing plant.”@en
DL2: OID_02 $\equiv$ “A fruit of the tree Prunus armeniaca.”@en
DL3: OID_02 $\sqsubseteq$ OID_01

From these, the following axioms can be inferred:

DL4: OID_02 $\sqsubseteq$ “A mature ovary of a seed-bearing plant.”@en
DL5: “A fruit of the tree Prunus armeniaca.”@en $\sqsubseteq$ “A mature ovary of a seed-bearing plant.”@en

DL4 can be reverse translated as:

OS4: OID_02 | Analytic | has_NC | “A mature ovary of a seed-bearing plant.”@en

However, reverse translation cannot be applied to DL5, as it would result in a statement without an OID as its *subject (*e.g.,
*“A fruit of the tree Prunus armeniaca”@en | Analytic | has_NC | “A mature ovary of a seed-bearing plant.”@en*, which is not a valid OID statement).

*4.3. Meaning specification of an ontological unit*

The “meaning specification” of an ontological unit aims to specify the intensional relation identified by the OID associated with the OU by incorporating the relevant OID statements that characterize that intensional relation. As outlined earlier (cf. 3.2), these relevant statements, referred to as “meaning-specifying OID statements,” are defined as follows:

*Meaning-specifying OID statement=def. An OID statement that is analytic, not logically reducible and whose condition type specifies a necessary (that might be necessary and sufficient) condition.*

These statements can be asserted directly in an OU or they can be inferred during the computation of the meaning specification. To effectively illustrate the meaning specification derived from the asserted meaning-specifying OID statements, we will examine the following set of assertive OID statements relevant to the domain represented in Figure 3.

(EX$_1$) OID_01 | Analytic | has_NSC | “A mature ovary of a seed-bearing plant.”@en

($EX_2$) OID_01 | Analytic | has_NC | ¬OID_99
($EX_3$) OID_01 | Analytic | has_SC | “Apple”@en
($EX_4$) OID_02 | Analytic | has_NSC | “A fruit of the tree Prunus armeniaca.”@en
($EX_5$) OID_02 | Analytic | has_NC | OID_01
($EX_6$) OID_02 | Synthetic | has_NC | “Contains vitamin A.”@en
($EX_7$) OID_02 | Analytic | has_NC | OID_99 ⊔ ¬OID_99
($EX_8$) OID_03 | Analytic | has_NC | OID_01
($EX_9$) OID_03 | Analytic | has_NC | “A tropical fruit.”@en
($EX_{10}$) OID_99 | Analytic | has_NC | ¬OID_01
($EX_{11}$) OID_99 | Analytic | has_NSC | “A rock”@en

Our focus will be on a specific OU, encapsulating the intensional relation identified by “OID_02” and referred to as “$OU_{OID_02}$.” In this case, the author aims to represent and communicate about their intensional relation of what is an apricot. For that purpose, they create an OID (“OID_02”) to identify this intensional relation, along with a set of OID statements: {$EX_4$, $EX_5$, $EX_6$, $EX_7$}.

Among all these statements $EX_1$-$EX_{11}$, $EX_3$ expresses a sufficient condition, and $EX_6$ is a synthetic statement. While they may help readers understand the intensional relation identified by the OID, they are not part of an OU’s meaning specification: synthetic statements may be empirically falsified, and sufficient condition statements do not make a general claim about their subject. Analytic OID statements for $OU_{OID_02}$ are the following:

($EX_4$) OID_02 | Analytic | has_NSC | “A fruit of the tree Prunus armeniaca.”@en
($EX_5$) OID_02 | Analytic | has_NC | OID_01
($EX_7$) OID_02 | Analytic | has_NC | OID_99 ⊔ ¬OID_99

$EX_7$ is a logically reducible statement as its characterization (“OID_99 ⊔ ¬OID_99”) can be reduced to “⊤” and is therefore excluded from $OU_{OID_02}$’s meaning specification. The first two statements ($EX_4$, $EX_5$) specify OID statements and constitute the asserted meaning specification of this OU that is defined as:

*Asserted meaning specification of an ontological unit=def. The collection of all asserted meaning-specifying OID statements of this ontological unit.*

However, these statements in our example are insufficient to comprehensively constitute the OU’s meaning specification, as their characterization mentions other OID (e.g., OID_01 in $EX_5$). In order to do so, several steps must be taken. First, one must recursively include all meaning-specifying OID statements whose OID are referenced in the original statements’ characterization. These statements form the analytic theory of an OU that is defined as follows:

*Analytic theory of an ontological unit=def. The collection of meaning-*specifying OID statements *that includes the asserted meaning specification of this ontological unit, along with the asserted meaning specification of all OUs which are recursively mentioned within it.*

To note, the analytic theory for a given OU may include OID statements belonging to other OUs (e.g., $EX_1$ and $EX_2$ belong to OID_01 OU). In our example, the analytic theory of $OU_{OID_02}$ is as follows:

($EX_1$) OID_01 | Analytic | has_NSC | "A mature ovary of a seed-bearing plant."@en
($EX_2$) OID_01 | Analytic | has_NC | ¬OID_99
($EX_4$) OID_02 | Analytic | has_NSC | "A fruit of the tree Prunus armeniaca."@en
($EX_5$) OID_02 | Analytic | has_NC | OID_01
($EX_{10}$) OID_99 | Analytic | has_NC | ¬OID_01
($EX_{11}$) OID_99 | Analytic | has_NSC | "A rock"@en

Second, leveraging the underlying logic described above, this analytic theory, including statements that have a natural language characterization, can be translated into DL, enabling the derivation of new DL axioms. These axioms can then be reverse translated into inferred meaning-specifying OID statements. In our example, two new inferred meaning-specifying OID statements about OID_02 are generated for the OU:

($EX_1^*$) OID_02 | Analytic | has_NC | "A mature ovary of a seed-bearing plant."@en
($EX_2^*$) OID_02 | Analytic | has_NC | ¬OID_99

As a reminder, "A mature ovary of a seed-bearing plant."@en belongs here to the nonlogical vocabulary of the theory: more specifically, it is the name of a class. These inferred meaning-specifying OID statements are added to the analytic theory of OID_02 OU to constitute its deductive closure. Then, all asserted and inferred meaning-specifying OID statements whose subject are "OID_02" are extracted. In our example, these meaning-specifying OID statements are the following statements:

($EX_1^*$) OID_02 | Analytic | has_NC | "A mature ovary of a seed-bearing plant."@en
($EX_2^*$) OID_02 | Analytic | has_NC | ¬OID_99
($EX_4$) OID_02 | Analytic | has_NSC | "A fruit of the tree Prunus armeniaca."@en
($EX_5$) OID_02 | Analytic | has_NC | OID_01

These meaning-specifying OID statements constitute the meaning specification of an ontological unit (OID_02 OU in our example), which is defined as follows:

*Meaning specification of an ontological unit = def. The collection of all asserted meaning-specifying OID statements from the OU's analytical theory, along with the meaning-specifying OID statements logically deduced from this analytical theory, that have as subject the OID associated with the OU.*

Consequently, the meaning specification of an OU cannot be determined entirely independently from other OUs. However, this interdependence enables us to identify the external statements the meaning specification depends on, and to evaluate the extent to which inferred meaning-specifying OID statements derived from the combination of multiple OUs may belong to the meaning specification of each.

### *4.4. Definition of an ontological unit*

To summarize the approach outlined above, consider an author aiming to represent and communicate about something (e.g. apricots). The author wants to communicate their own intensional relation (e.g. $\rho_{\text{apricot}}$) which is a function that assigns to each possible world a specific extension (ex. $\rho_{\text{apricot}}$ maps $w_1$ to all apricots in $w_1$, $w_2$ to all apricots in $w_2$, etc.). The author uses an OID to identify their intensional relation (ex. “OID_02”). This association represents the author’s ontological commitment, linking the OID to extensional relational structures across all possible worlds that constitute the author’s set of intended models.

To communicate this ontological commitment to putative[6] readers, the author creates an ontological unit (ex. $OU_{OID_02}$), which includes a set S of statements with among them meaning-specifying OID statements (ex. $EX_4$ and $EX_5$ in the earlier example). From S, a meaning specification can be derived (ex. $EX_1$*, $EX_2$*, $EX_4$, $EX_5$ from the same example).

As the reader works to understand the OU’s meaning specification, they develop their own ontological commitment and therefore have their own mapping between the OID and extensional relational structures across all possible worlds. These constitute the reader’s set of intended models, referred to as “understood models” to distinguish them from the author’s intended models.

The ultimate goal of the OU’s author is that for each world *w*, the corresponding extensional relational structure of the reader’s understood model is identical to the extensional relational structure of their intended model in that world. This implies that authors and readers have identical intensional relations and consequently identical ontological commitments. However, this identity cannot be verified in practice, as there might be a very large number of possible worlds. Thus, the author’s objective is to approximate this identity rather than achieve it perfectly. Resulting from these considerations, the following definition of an ontological unit is proposed:

*Ontological unit=def. An information content entity that includes a collection of meaning-specifying OID statements for an OID of interest with the intent of allowing a reader to generate the corresponding intension through the derivation of a meaning specification; it also include metadata and may include non meaning-specifying OID statements, such as synthetic statements or statements expressing a sufficient condition, which enable accurate contextualization of the information it encapsulates.*

## 5. Discussion

This work addresses the challenge of accurately representing and communicating the meaning of an ontological term using ontological units. This work is situated within the context of ontology interoperability, where the most common strategy for addressing interoperability challenges is ontology alignment, which entails establishing semantic correspondences between ontological terms drawn from distinct ontologies (Osman et al., 2021). To support the exchange and reuse of these alignments, a standardized mapping format has been proposed (Matentzoglu et al., 2022).

Although ontologies alignment can be partially automated, it remains labor-intensive and typically requires human curation to accurately interpret the intended meanings of both source and target terms. This process can also lead to semantic ambiguities as aligned terms are not always strictly equivalent. Moreover, alignment is generally performed on a point-to-point basis which increases the number of mappings required as the number of ontologies grows and may hinder the long-term sustainability of the mappings. Our approach does not address ontology alignment as it involves ontological units and not ontologies themselves.

---

[6] We consider only for now readers such as envisioned by the author but recognize that the actual readers may differ from this expectation.

Additionally, our method differs fundamentally from ontology modularization techniques (Cuenca Grau, Horrocks, Kazakov, et al., 2008), such as MIREOT (Courtot et al., 2011) and Locality-Based Modules (LBM) (Del Vescovo et al., 2012) as these techniques aim to segment an ontology, while our method starts from already segmented elements, namely OUs, enabling the generation of ontologies.

Nevertheless, by specifying the meaning of terms in a precise and structured manner, ontological units have the potential to facilitate the alignment and modularization of the ontologies derived from them.

We methodologically followed the conception proposed by Guarino et al. (2009) by equating the meaning of an ontological unit with an intensional relation. This approach has faced criticism, notably from Neuhaus (Neuhaus, 2017), who argues that defining conceptualization on the basis of intensional relations fails to capture: 1) the complexity of human conceptualizations, 2) the possibility of incomplete knowledge, and 3) the reality of the conceptual division of labor. While these critiques are relevant, addressing them in totality lies beyond the scope of this work. Here, intensional relations are employed as practical tools for representing and operationalizing meaning, without assuming that it has a cognitive nature or its relation to the term "conceptualization". Given the influence of Guarino et al.'s conception of an ontology, they provide a useful tool to present a switch from an ontology-centered conception to a conception centered around OU.

Examples in this work are restricted to intensional relations represented in DL as concepts. However, we are currently exploring the application of this approach to other DL structures, including roles and datatypes. Preliminary analysis suggests that the proposed formalism is broadly applicable. Indeed, many role characteristics can be expressed by subsumption axioms. For example, the axiom "$\exists \text{OID_06}.\top \sqsubseteq \text{OID_07}$" asserts that OID_07 is a domain of the role "OID_06", which can be represented e.g. as a sufficient condition on OID_07 within our approach. However, not all role characteristics can be expressed in this manner (consider asymmetry).

Note that even general class axioms can be represented in our approach. For example, an axiom such as "$\exists \text{OID_28}.\top \sqsubseteq \exists \text{OID_10}.\text{OID_11}$" cannot be directly translated in an OID statement as neither "$\exists \text{OID_28}.\top$" nor "$\exists \text{OID_10}.\text{OID_11}$" can serve as the subject of a valid OID statement. However, it can be represented by creating a dedicated ontological unit with a specific OID that points at one of the intensional relations evocated by the general class axiom (in the example above, creating an OID equivalent to $\exists \text{OID_28}.\top$ or to $\exists \text{OID_10}.\text{OID_11}$) and expressing the general class axiom as a necessary or sufficient condition statement on this OID.

From a formal perspective, approaches such as axiomatic systems are not addressed in this work. The proposed method adopts a top-down approach to meaning characterization that is justified by the hierarchical nature of meaning specification and aligns well with implementation in Description Logic (DL). Our primary aim is to develop an approach that is both computationally feasible and relevant to key ontology applications. Future work will explore alternative formalisms, such as first-order logic, particularly given its increasing integration within OWL 2 (Flügel et al., 2024). Similarly, this approach is oriented toward the representation of scientific domains characterized by clearly delimited entities. More advanced forms of representation, such as those involving conceptual blending (Righetti et al., 2021), have not yet been explored. Moreover, the proposed characterization of meaning is grounded in strict formal constraints such as necessary and sufficient conditions and is likely to limit the flexibility required for such methods.

The intensional relations introduced in this paper are grounded in possible worlds semantics, as they have an extension in each possible world. However, the approach proposed by Guarino et al. assumes a fixed domain, meaning that the elements composing these extensions exist in every possible world. The use of variable-domain semantics (Priest, 2008), which allows elements of the universe of discourse to exist in some worlds but not in others, could be considered in future works, as it could enable the representation of potential or fictional entities.

Another important feature of ontological units is the improved characterization of statements, whether in formal or natural language. By explicitly distinguishing between analytic and synthetic statements, our framework enables authors to differentiate the defining features of a unit—captured in analytic statements—from non-defining features that offer heuristics to aid user understanding. Additionally, the clarification of the type of condition expressed allows for a more precise characterization of natural language statements, such as a necessary and sufficient condition, compared to the above-mentioned annotation property IAO_0000115, which is used irrespective of the condition expressed.

The integration of natural language statements with the formal statements has already been proposed, notably by Neuhaus in the form of an "annotated logic theory" (Neuhaus, 2018). The author highlights the importance of assertive annotations for capturing information not represented by logical axioms and clarifying potential ambiguities in vocabulary and axioms. However, since the interpretation of these annotations are carried out in parallel with that of the logical axioms, inconsistencies may arise between the logical theory and the assertive annotations in an annotated logical theory. Incorporating elements of natural language into the logical vocabulary and providing a unified framework for their interpretation helps mitigate this issue.

The proposed formal definition of an ontological unit's meaning aims to enhance the semantic robustness of ontological terms in view of the phenomena of meaning holism and reference indeterminacy. In our context, it is worth noting that the classical phenomenon of indeterminacy of reference concerns here the question of which intensional relation is identified by an OID, by each agent using this OID: since intensional relation is a good proxy for meaning, we have here a phenomenon of indeterminacy of meaning rather than indeterminacy of reference. Similarly, the classical phenomenon of meaning holism is here reflected in the fact that the meaning specification of an ontological unit depends on the meaning specification of other ontological units (namely, those that are associated with OIDs recursively mentioned by the statements in the meaning specification). Therefore, the corresponding phenomenon should rather be named here "meaning *specification* holism".

Ontologies serve diverse purposes, ranging from hierarchically structured controlled vocabularies—where annotations and comments play a vital role in providing a standardized set of community-approved terms—to information systems that leverage their logical foundations to answer data queries. Our approach aims to provide added value for these purposes. First, an ontological unit functions as a repository of meaning postulates, enhancing maintainability and usability as a reference resource and would allow a better compartmentalization of meaning and may serve as a tool for implementing meaning dependencies (Neuhaus et al., 2025) (Figure 5).

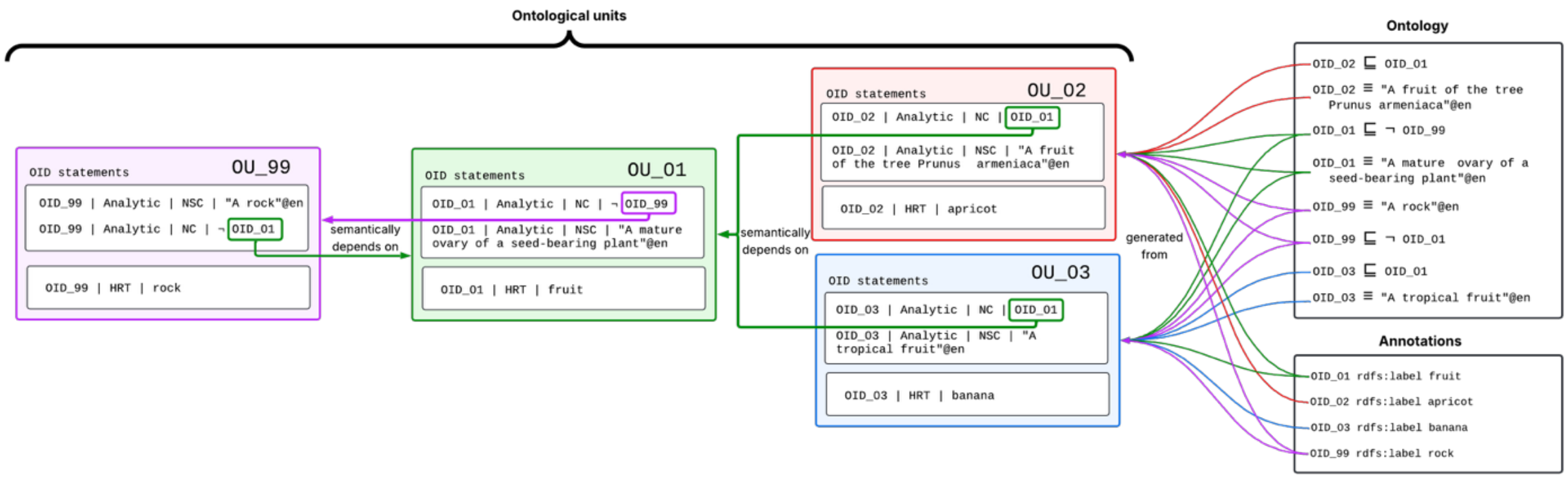

Figure 5 - Generation of an ontology from a set of ontological units under the proposed framework. This enables tracing each ontology axiom to its originating ontological unit and supports the evaluation of inconsistencies across ontological units.

This positions it as a foundational element conceptually upstream of ontological engineering. Second, generating OWL ontologies from OID statements coming from multiple ontological units would ease the creation and maintenance of ontologies tailored to specific information systems. This would ensure that ontological units can be effectively integrated at the onset of ontology development and deployment.

## 6. Conclusion and future works

While this work lay the theoretical groundwork for a formal specification of an ontological term's meaning via the use of ontological unit, it is rooted in considerations arising from the practical application of ontologies, notably for term reuse and version control.

Reusing ontologies (i.e., importing parts of one ontology into another) is a fundamental practice in ontology design, but it has become increasingly challenging, especially when imports are layered or chained. The fundamental principle of OUs is to enable fine-grained control over the meaning of terms, at the level of individual terms, rather than entire ontologies. By employing OUs, it becomes possible to automatically verify whether the association of different OUs alters their specified meanings, thereby offering an indication of potential meaning distortion, irrespective of the contexts in which these OUs were originally created.

Version control has been widely discussed in the context of ontologies. Neuhaus, for instance, highlights the importance of versions in ontology definition ("An ontology of a domain is a document that is realized by a network of ontology versions about the domain." in Neuhaus, 2018). Since ontologies evolve over time, different versions of the same ontology can be associated with different sets of formulas.

Although recent efforts have been made to formalize the description of changes (Hegde et al., 2024), version documentation in practice is often limited to modifying the version date in the ontology's metadata and applying it to the entire ontology without identifying which specific classes have been modified. This becomes even more problematic when users import only a subset of an ontology, which may include thousands of classes, leaving it unclear which ones are affected by the update.

At the ontological term level, versioning often amounts to inactivating a term by making it obsolete, without clear guidelines for taking this decision. While changes to axioms can be objectively evaluated using a reasoner, modifications to natural language definitions are far more challenging to assess, particularly when determining whether creating a new term is warranted or whether modifying the definition of the existing one is acceptable. In addition, the decision to deprecate certain terms also introduces unique challenges for users who rely on those terms.

Versioning at the level of individual ontological units would offer more fine-grained control of OID statements. Modifying a meaning-specifying OID statement would necessitate the creation of a new ontological unit, as such a change, despite potentially preserving the same intensional relation from the author's perspective, may introduce a new ontological commitment for users. On the other hand, modifications to other types of statements such as human-readable pointers or synthetic OID statements, which do not change the meaning of the associated OID, could instead warrant the release of a new version of the associated OU. Both of these issues will be examined in greater detail in future work.

## 7. Acknowldegment

We thank both reviewers for their time and effort in evaluating the manuscript as well as Fabian Neuhaus for worthwhile discussions. We are sincerely grateful for their insightful comments and constructive suggestions, which have significantly contributed to improving the quality of the work.